\documentclass[fleqn,usenatbib]{mnras}

\usepackage{newtxtext,newtxmath}

\usepackage[T1]{fontenc}
\usepackage{ae,aecompl}

\usepackage{graphicx}	
\usepackage{amsmath}	
\usepackage{amssymb}	

\title[Frequencies of stellar models with 3D-atmospheres]{Theoretical oscillation frequencies for solar-type dwarfs from stellar models with $\langle 3\mathrm{D} \rangle$-atmospheres}

\author[A. C. S. J\o rgensen et al.]{
\and
Andreas Christ S{\o}lvsten J{\o}rgensen$^{1}$ \thanks{E-mail: acsj@mpa-garching.mpg.de},
Achim Weiss$^{1}$,
Jakob R{\o}rsted Mosumgaard$^{2}$,
\and
Victor Silva Aguirre$^{2}$,
and
Christian Lundsgaard Sahlholdt$^{2}$
\\
$^{1}$Max Planck Institut f\"ur Astrophysik, Karl-Schwarzschild-Strasse 1, 85748 Garching, Germany\\
$^{2}$Stellar Astrophysics Centre, Department of Physics and Astronomy, Aarhus University, Ny Munkegade 120, DK-8000 Aarhus C, Denmark\\
}

\date{Accepted 08.2017. Received YYY; in original form ZZZ}

\pubyear{2017}

\begin{document}
\label{firstpage}
\pagerange{\pageref{firstpage}--\pageref{lastpage}}
\maketitle

\begin{abstract}

We present a new method for replacing the outermost layers of stellar models with interpolated atmospheres based on results from 3D simulations, in order to correct for structural inadequacies of these layers. This replacement is known as patching. Tests, based on 3D atmospheres from three different codes and interior models with different input physics, are performed. Using solar models, we investigate how different patching criteria affect the eigenfrequencies. These criteria include the depth, at which the replacement is performed, the quantity, on which the replacement is based, and the mismatch in $T_\mathrm{eff}$ and $\log g$ between the un-patched model and patched 3D atmosphere. We find the eigenfrequencies to be unaltered by the patching depth deep within the adiabatic region, while changing the patching quantity or the employed atmosphere grid leads to frequency shifts that may exceed $1\,\mu\mathrm{Hz}$. Likewise, the eigenfrequencies are sensitive to mismatches in $T_\mathrm{eff}$ or $\log g$. A thorough investigation of the accuracy of a new scheme, for interpolating mean 3D stratifications within the atmosphere grids, is furthermore performed. Throughout large parts of the atmosphere grids, our interpolation scheme yields sufficiently accurate results for the purpose of asteroseismology. We apply our procedure in asteroseismic analyses of four \textit{Kepler} stars and draw the same conclusions as in the solar case: Correcting for structural deficiencies lowers the eigenfrequencies, this correction is slightly sensitive to the patching criteria, and the remaining frequency discrepancy between models and observations is less frequency-dependent. Our work shows the applicability and relevance of patching in asteroseismology.
\end{abstract}




\section{Introduction}

Over the last decades, asteroseismology, i.e. the study of stellar oscillations, has led to a unique insight into the structure of stars and contributed to exoplanet research. This has been been accomplished based on observational constraints from the CoRoT \citep{Baglin2009} and the \textit{Kepler} \citep{Borucki2010} space missions, as stellar parameters can be extracted from asteroseismic measurements, i.e. from a detailed comparison with the predictions of stellar structure models \citep{jcd2010, Aguirre2013, Lebreton2014}.

However, current one-dimensional stellar evolution codes give an inadequate representation of the outermost layers of stars with convective envelopes, due to the parametrization needed to describe turbulent convection, when dealing with the long time-scales of nuclear burning. For the outermost superadiabatic layers, the inadequate depiction results in theoretical oscillation frequencies that systematically underestimate the frequencies obtained from asteroseismic measurements, at high frequencies. These frequency shifts are collectively known as surface effects. In order to correct for these systematic errors in the model frequencies, empirical surface-effect corrections are often employed in asteroseismic analysis. Most authors employ the empirical correction introduced by \citet{Kjeldsen2008}, including free parameters that are calibrated to fit solar observations.  A recently published study by \citet{Sonoi2015}, based on 3D hydrodynamical simulations, has meanwhile found this empirical correction not to be applicable across the Hertzsprung--Russell (HR) diagram and suggests an alternative Lorentzian formulation. 

\citet{Ball2014} have likewise tested different parametrizations of surface corrections, including the empirical power-law fit by \cite{Kjeldsen2008} as well as parametrizations with a cubic term proportional to $\nu^3/\mathcal{I}$. Here $\nu$ and $\mathcal{I}$ denote the eigenfrequencies and the mode inertia, respectively. Using degraded solar data and observations of HD~52265, they find the cubic term to dominate the surface correction and the cubic correction to perform better than the power-law fit by \citet{Kjeldsen2008}.

To avoid the necessity of empirical relations to correct for surface-effects, different authors, including \citet{Schattl1997} and \citet{Rosenthal1999}, have improved the  surface layers of 1D models, based on 2D or 3D hydrodynamic atmosphere simulations. \citet{Piau2014} later suggested to fully replace the outermost layers by the mean stratification of these simulations, to which we will refer as $\langle 3\mathrm{D} \rangle$-atmospheres. This approach, known as patching, has been shown to lead to reliable results in the case of solar models. It has the advantage that it allows for the study of different physical parameters, such as the magnetic field strength, as was done in an analysis performed by \citet{Magic2016b}. Furthermore, following this scheme, both \citet{Sonoi2015} and \citet{Ball2016} have presented patched models in different regions of the HR diagram and obtained better agreement with observed solar pulsation frequencies.

However, as 3D hydrodynamic atmosphere simulations are computationally cumbersome, only patched models, based on precomputed 3D simulations, are available alternatives. While the patching procedure has thus been used to calibrate relations for surface corrections, it has only been applied directly in a very limited set of cases, including the Sun. In other words, using the sketched replacement of the outermost layers in asteroseismic analyses generally requires a reliable interpolation scheme, based on existing grids of 3D atmosphere simulations, to avoid computationally intensive calculations \citep[e.g.][]{Magic2016a}.

The aim of this paper is to present a new and more robust patching method, including an alternative interpolation scheme. Moreover, we aim to shed light on the impact of stellar patching on asteroseismic analyses, based on solar models, and present first tests, based on \textit{Kepler} data. 

These theoretical oscillation frequencies are not expected to perfectly match observations, as the patching procedure only corrects for structural effects. More specifically, analyses, based on patched models, are expected to slightly underestimate the observed frequencies, and modal effects must be taken into account, in order to correct for the remaining frequency difference. Modal effects include non-adiabatic energetics and contributions from the turbulent pressure to the oscillation equations. According to \citet{Houdek2017}, modal effects counteract structural effects, reducing the purely structural frequency correction, yielding results that are in very good agreement with observations. Including modal effects is, however, beyond the scope of this paper. Nevertheless, our analysis illustrates the potential of stellar patching, in connection with asteroseismology. 

The general features of our patching procedure are described in Section~\ref{section:methods}. Having established this procedure, we present a handful of patched solar models in Section~\ref{sec:solar}, as this allows for a straight-forward comparison with the results published by \cite{Ball2016}. Moreover, we use these models to investigate how different assumptions and choices affect the seismic results, as these questions have not yet been fully addressed by other authors. These choices include the quantities, on which to base the patching procedure, as well as the matchmaking between the interior and atmosphere models.

In order to test our interpolation method, we have reproduced models throughout existing atmosphere grids --- the main results hereof are briefly discussed in Section~\ref{section:methods}. Furthermore, to investigate how errors that are introduced by the interpolation affect the model frequencies, we present an asteroseismic analysis of patched solar models, based on interpolated atmospheres, in Section~\ref{sec:solar}. In this connection, we also investigate how mismatches in the effective temperature and gravitational acceleration between the un-patched model and the patched atmosphere affect the model frequencies.

Finally, we present an asteroseismic analysis of four stars in the \textit{Kepler} LEGACY dwarfs sample~I \citep{Lund2017, Aguirre2017} and KAGES sample \citep{Davies2016, Aguirre2015}, based on patched stellar models. The results of this analysis can be found in Section~\ref{sec:KeplerPatch}. Section~\ref{sec:Conclusion} contains our main conclusions.

\section{Constructing Patched Models} \label{section:methods}

We construct patched stellar models (PMs) by replacing the outermost stratification of un-patched models (UPMs) by the mean stratification of 3D stellar atmospheres. In accordance with other authors, we perform the replacement at a given evolutionary stage, and we hence refer to our procedure as post-evolutionary patching. The previous evolution up to this point is therefore computed without replacing the outermost layers.

The employed 3D simulations are all of the \textit{box-in-a-star} type, covering a representative volume of the surface layers, and the mean stratifications ($\langle 3\mathrm{D}\rangle$) are both spatial averages over horizontal layers and temporal averages.

To conserve hydrostatic equilibrium, the spatial average is taken over layers of constant geometrical depth. As shown in \citet{Magic2016b}, using other averages, such as averages taken over Rosseland optical depth, violates hydrostatic equilibrium and consequently affects the seismic results. According to \cite{Magic2016b}, the mean stratifications are, however, not affected by whether the temporal average is taken over layers of constant depth or over horizontally averaged layers of constant column density.

\subsection{The Interior Models} \label{subsec:interior}

When constructing post-evolutionary PMs, it is assumed that the inadequate treatment of the outermost layers by the 1D stellar evolution code has a negligible effect on stellar evolutionary tracks. This assumption also underlies the application of the mentioned empirical relations for surface corrections, but it is unclear, to which extent it holds true. A more rigorous approach would thus be to patch $\langle 3 \mathrm{D} \rangle$-atmospheres throughout the stellar evolution, to take the results of 3D simulations consistently into account, along the evolutionary path. This, however, lies beyond the scope of the present work but may be pursued in a subsequent paper.

In order to construct UPMs, whose input physics closely match the assumptions made in the 3D simulations, we use the $T(\tau)$ relation and calibrated $\alpha_{\mathrm{MLT}}$ that were introduced by \citet{Trampedach2014a} and \citet{Trampedach2014b}, based on a grid of $\langle 3\mathrm{D} \rangle$-atmospheres of solar-like composition. Both the $T(\tau)$ relation and calibrated $\alpha_{\mathrm{MLT}}$ have been implemented in the \textbf{GA}rching \textbf{ST}ellar \textbf{E}volution \textbf{C}ode, \textsc{garstec} \citep{Weiss2008}, as described by \citet{Mosumgaard2016}. In this paper, we therefore restrict ourselves to models and objects at solar-like composition. The input physics are further specified in Section~\ref{sec:KeplerPatch}.

These \textsc{garstec} models are employed when dealing with the \textit{Kepler} stars in Section~\ref{sec:KeplerPatch}. For the sake of comparison, we also construct PMs, using \textsc{garstec} models with different input physics. In the case of the analysis of patched solar models presented in Section~\ref{sec:solar}, however, we have used un-patched standard solar models, based on the \textbf{A}arhus \textbf{ST}ellar \textbf{E}volution \textbf{C}ode, \textsc{astec} \citep{jcd2008a}, to facilitate an easy comparison with the results published by \citet{Ball2016}.

\subsection{The Atmosphere Grids} \label{subsec:grids}

For our purposes, we had access to two grids of 3D radiation-coupled hydrodynamic simulations of stellar convection. Firstly, we use a grid by \citet{Trampedach2013}, to which we will refer as the Trampedach-grid, containing 37 atmospheres, all at solar metallicity. Secondly, we use a grid by \citet{Magic2013a}, to which we will refer as the Stagger-grid, consisting of 206 models with metallicities, ranging from $\mathrm{[Fe/H]=-4.0}$ to $\mathrm{[Fe/H]=0.5}$, of which 29 models are at solar metallicity. We exclude one of the models at solar-composition, as the computation does not seem to be fully relaxed.

As the Trampedach-grid and Stagger-grid are based on the successors of the \citet{Stein1998} code, the numerical schemes used in the respective computation of the two grids are rather similar. Furthermore, both grids use the so-called Mihalas-Hummer-D\"appen equation of state \citep[MHD-EOS,][]{Hummer1988}. The Trampedach-grid is based on abundances found in \citet{Trampedach2013} and line opacities found in \citet{Kurucz1992a} and \citet{Kurucz1992b}, while the Stagger-grid employs the composition listed in \citet{Asplund2009} and MARCS line opacities found in \citet{Gustafsson2008}. The two grids are two distinct groups of models, relying on somewhat different input parameters, as well as technical and numerical details.

\subsection{The Patching Procedure} \label{subsec:PatchingProcedure}

\begin{figure}
\centering
\includegraphics[width=\linewidth]{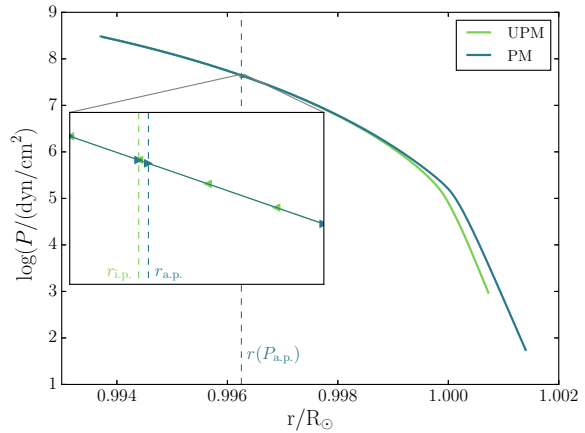}
\caption{Pressure stratification of a patched model (PM) and an un-patched model (UPM), based on Model S and the solar Stagger-grid atmosphere. The total pressure, $P$, is the patching quantity, entering the selection of the patching points and corresponding radii. The zoom-in includes the interior patching point (i.p.), i.e. the outermost mesh point of the interior model, and the atmosphere patching point (a.p.), i.e. the innermost mesh point of the patched atmosphere, as well as the associated radii: $r_{i.p.}$ and $r_{a.p.}=r(P_{a.p.})$.
}
\label{fig:IllustParam}
\end{figure}

When constructing a patched model (PM), we discard all mesh points in the un-patched model (UPM) beyond a certain interior patching point (i.p.). Likewise, we discard all mesh points below the so-called atmosphere patching point (a.p.), in the $\langle 3\mathrm{D} \rangle$-atmosphere. We then replace the stellar structure of the UPM beyond the interior patching point with the remaining mesh points from the $\langle 3\mathrm{D} \rangle$-atmosphere. Fig.~\ref{fig:IllustParam} illustrates the corresponding change in the stellar structure and the nomenclature that is introduced in this section.

There are different viable approaches one may take, when selecting the patching points. One may preselect the atmosphere patching point, at, say, a given pressure or depth. The patch is then performed, at the distance $r(\beta_\mathrm{a.p.})$ from the stellar centre of the UPM, at which a given patching quantity, $\beta$, takes the same value in the UPM as at the atmosphere patching point. Since a mesh point, matching the value of the patching quantity at the atmosphere patching point, may not exist in the UPM, this approach requires interpolation in the UPM. As discussed in Section~\ref{sec:solar}, suitable choices for the patching quantity, $\beta$, include the temperature $T$, the total pressure ($P$), the density ($\rho$), and the first adiabatic index ($\Gamma_1$). 

Instead of preselecting the atmosphere patching point, one may preselect the interior patching point. Furthermore, one may define patching criteria, based on two or more patching quantities. Involving more than one patching quantities, one may also select the set of patching points that overall gives the best match between the $\langle 3\mathrm{D} \rangle$-atmosphere and the interior model, for these patching quantities. For instance, one may perform the patch at that point in the $(P,\rho)$-plane, at which $\rho(P)$ predicted by the $\langle 3\mathrm{D} \rangle$-atmosphere lies closest to $\rho(P)$ predicted by the UPM. 

When involving several patching quantities, the values of the patching quantities, at the atmosphere patching point, are most probably not all matched by the UPM at the same distance from the stellar centre. In other words, each patching quantity assigns a different distance from the stellar centre to the atmosphere patching point, and one therefore has to define a suitable weighted mean of these distances. When finding the best match in the $(P,\rho)$-plane, we have thus weighted $r(\rho_\mathrm{a.p.})$ and $r(P_\mathrm{a.p.})$ by the respective discrepancies of these quantities between the interior and the atmosphere patching point.

In this paper, we follow several of the approaches described above, in order to evaluate the influence of different patching criteria, or in order to facilitate an easy comparison with \citet{Ball2016}. We elaborate on this in Section~\ref{sec:solar}. 

None of the listed patching methods explicitly ensures the conservation of hydrostatic equilibrium, at the patching points. We therefore also construct models, for which we determine the distance, $r_\mathrm{a.p.}$, of the atmosphere patching point from the stellar centre of the UPM, in such a way as to ensure hydrostatic equilibrium to first order:
\begin{equation}
r_\mathrm{a.p.}= r_\mathrm{i.p.}-\frac{2 \left(P_{\mathrm{a.p.}}-P_{\mathrm{i.p.}}\right) r_\mathrm{\mathrm{i.p.}}^2}{G\cdot m(r_{\mathrm{i.p.}}) \left(\rho_{\mathrm{a.p.}}+\rho_{\mathrm{i.p.}}\right)}. \label{eq:HSEfit}
\end{equation}
Here the index i.p. refers to the interior patching point, i.e. the outermost mesh point in the UPM, for which the distance to the centre is smaller than $r(\beta_\mathrm{a.p.})$. $G$ denotes the gravitational constant, and $m(r)$ denotes the mass interior to $r$.

We finally note that 3D hydrodynamic simulations are computed in the plane-parallel approximation, assuming constant surface gravity, and we hence adjust $\log g$, at the atmosphere patching point, based on the interior model, and correct for sphericity, when constructing PMs. This is done iteratively: We compute the difference between the assumed constant gravitational acceleration of the $\langle 3\mathrm{D} \rangle$-atmospheres and the actual gravitational acceleration at the associated distance from the stellar centre of the PMs, starting at the lowermost mesh point. We then use this difference in gravitational acceleration to correct the distance, $\mathrm{d} z$, between the considered mesh point and its next neighbour in the patched atmosphere, based on hydrostatic equilibrium:
\begin{equation}
\frac{\mathrm{d}P}{\mathrm{d}z}=\rho g. \label{eq:hydroEq}
\end{equation}
For the shallow atmospheres of solar-like main sequence stars, we find this correction to have little influence.

\subsection{The Interpolation Scheme} \label{subsec:interpolation}

\begin{figure}
\centering
\includegraphics[width=\linewidth]{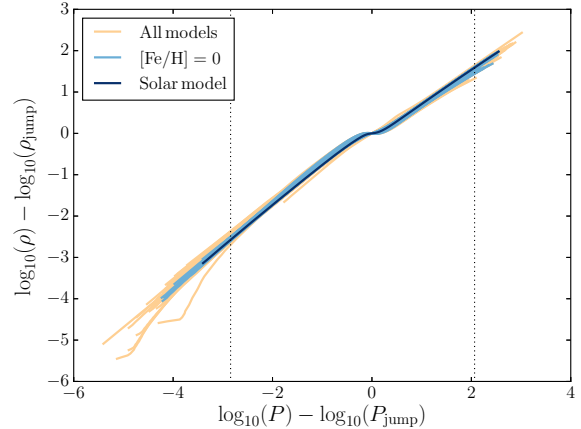}
\caption{The logarithm of the density as a function of the logarithm of the pressure after shifting the stratification by the position of the density jump for all considered 201 models in the Stagger-grid --- i.e. for $T_\mathrm{eff}$ ranging form $4000\,\mathrm{K}$ to $7000\,\mathrm{K}$, $\log g$ ranging from $1.5$ to $5.0$, and $\mathrm{[Fe/H]}$ ranging from $-4$ to $0.5$. The stratifications of the models at solar metallicity and the solar model have been highlighted. The dotted lines show the interpolation range for solar metallicity.
}
\label{fig:Prho}
\end{figure}

\begin{figure}
\centering
\includegraphics[width=\linewidth]{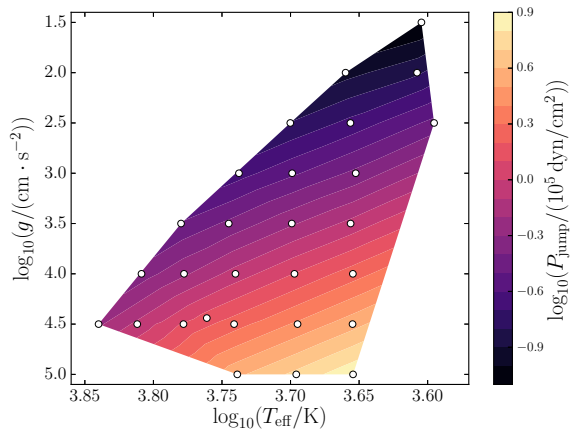}
\caption{The logarithm of the pressure, $\log P_\mathrm{Jump}$, corresponding to the minimum of $\partial \log \rho/\partial \log P$, based on the Stagger-grid atmospheres (white dots), at solar metallicity.
}
\label{fig:Pm}
\end{figure}

Constructing meaningful PMs for any given model or star, based on discrete atmosphere grids, often requires interpolation. To motivate our suggested interpolation scheme, we start out by noting that $\log \rho$ as a function of $\log P$ looks similar across the grids --- that is, comparing simulations from the same grid, across all parameters, as well as comparing models from different grids. Thus, $\log \rho$ behaves linearly at high and low pressure, while a jump occurs near the stellar surface, corresponding to a local minimum in $\partial \log \rho/\partial \log P$. Moreover, the inclination and intercept of the linear regions are similar for all models. Shifting $\log \rho$ as a function of $\log P$ by the position of the density jump thus results in nearly coinciding stratifications, as can be seen from Fig.~\ref{fig:Prho}. This is relevant, as the position of the density jump in the ($\log \rho,\log P$)-plane behaves rather linearly as a function of both $\log T_\mathrm{eff}$ and $\log g$, as can be seen from Fig.~\ref{fig:Pm}. In other words, the position of the density jump can reliably be computed by interpolation\footnote{It is worth noting that while $\log \rho$ as a function of $\log P$ is thus suitable for interpolation across the mapped ($T_\mathrm{eff},\log g$)-plane, $\log P$ as a function of $\log \rho$ is not. This is due to the density inversion that can take place in the envelopes of red giants, a phenomenon found by \citet{Schwarzschild1975}.}. 

To compute interpolated atmosphere models, we select a suitable subset of atmospheres, having the same metallicity as the atmosphere that we attempt to construct. We then shift the stratification of each of these atmospheres by the position of the respective density jump, as described above. The interpolation now takes place in the ($\log T_\mathrm{eff},\log g$)-plane, where we compute $\log \rho$ for fixed values of $\log P$. Thus, triangulating the grid points, we construct piece-wise cubic, continuously differentiable surfaces, to interpolate $\log \rho$, in the ($\log T_\mathrm{eff},\log g$)-plane, for each value of $\log P$. The obtained stratification is then shifted by the corresponding density jump that is likewise obtained by cubic interpolation in the ($\log T_\mathrm{eff},\log g$)-plane. Fig.~\ref{fig:Pm} shows the pressure at the density jump, for all models in the Stagger-grid, at solar metallicity. As can be seen from the figure, the pressure at the density jump evolves rather linearly as a function of both $\log T_\mathrm{eff}$ and $\log g$. The same holds true at different metallicities and for the corresponding density. In the case of the Trampedach-grid atmospheres, the position of the minimum in $\partial \log \rho/\partial \log P$ is a well-behaved function of $\log T_\mathrm{eff}$ and $\log g$ as well.

Having obtained $\rho(P)$, the depth $z$ below a given point can now easily be established, based on the requirement of hydrostatic equilibrium --- cf.~equation~(\ref{eq:hydroEq}). Furthermore, our interpolation scheme computes both $\log T$ and $\Gamma_1$ as a function of $\log P$, based on an interpolation in the ($\log T_\mathrm{eff},\log g$)-plane. Before interpolating, it is important to shift the stratification of both quantities as a function of pressure by the pressure and the respective value of the quantity, at the  density jump.

The derivatives needed for frequency calculations are determined, based on the established stratifications, using Akima spline interpolation \citep{Akima1970}. 
While cubic splines may show ripples in the neighbourhood of discontinuities, this piece-wise cubic and continuously differentiable sub-spline interpolation method yields a smooth transition, even when encountering abrupt changes in the derivatives. This feature is important near the patching points.

As pointed out by \citet{Magic2013b}, the temperature stratification of $\langle \mathrm{3D} \rangle$-atmospheres may severely deviate from their 1D counterparts. It thus stands to reason that the naive patch between an interior model and an atmosphere with the \textit{same} $T_\mathrm{eff}$ may not yield a good fit in $T(r)$, when using other quantities to determine the patch, and vice versa. When constructing PMs for the stars in the \textit{Kepler} field, we thus compute a large sample of interpolated atmospheres, requiring $T_\mathrm{eff}$ and $\log g$ to lie within three standard deviations of the observational mean. We then select that interpolated atmosphere that minimizes the sum of squared relative differences in $T$, $\rho$, and $\Gamma_1$, near the bottom of the interpolated atmosphere, weighted by $(1+(100\cdot \Delta \log g)^2)$, where $\Delta \log g$ is the difference in $\log g$ at the surface. By imposing this restriction on $\log g$, we prefer models, whose gravitational acceleration match the value of the UPM, as any mismatch has to be subsequently corrected for, in order to obtain physically meaningful PMs. The interpolation takes place at a pressure that is $10^{1.9}$ times higher than at the density jump, as this value lies close to the highest pressure that is common for all models in the grids, so that no extrapolation is required.

In short, when dealing with \textit{Kepler} stars (cf.~Section~\ref{sec:KeplerPatch}), we select the interpolated atmosphere that results in the smoothest stratifications. In the case of the Sun, on the other hand, we use the existing solar atmospheres in the grids.

\begin{figure}
\centering
\includegraphics[width=\linewidth]{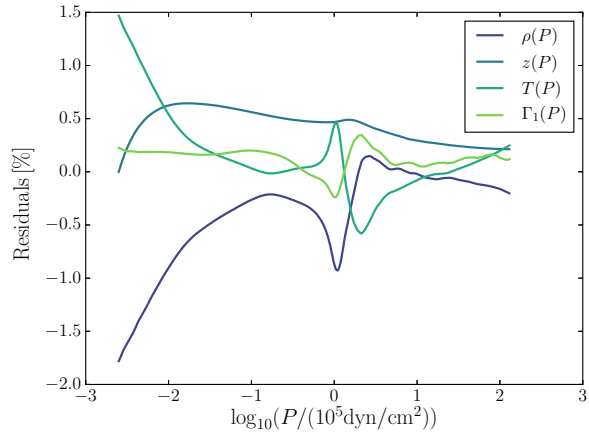}
\caption{Comparison between an interpolated solar model, based on the Stagger-grid, and the correct structure that was excluded from the grid prior to the interpolation. $T_\mathrm{eff}=5769\,\mathrm{K}$ and $\log g=4.44$. The figure shows the relative residuals for four quantities: the density ($\rho$), the adiabatic exponent ($\Gamma_1$), the temperature ($T$) and the depth ($z$) below the outermost point of the interpolated atmosphere as a function of pressure ($P$). 
}
\label{fig:residuals}
\end{figure}

To evaluate how accurately our interpolation scheme reproduces the correct structure across the ($\log T_\mathrm{eff},\log g$)-plane, we have excluded models from the grids and subsequently reconstructed these models. An example is given in Fig.~\ref{fig:residuals}, where we show the relative residuals between the solar model in the Stagger-grid and the corresponding interpolated atmosphere. As can be seen from the figure, all interpolated quantities are reproduced within $2\,\%$ of the expected values, at all pressures, ranging over five orders of magnitude. For the pressure range shown in Figure~\ref{fig:residuals}, all models in the atmosphere grid provide values for the pressure.

\begin{figure}
\centering
\includegraphics[width=\linewidth]{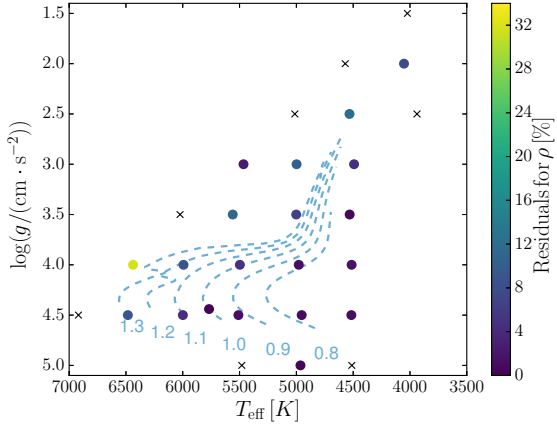}
\caption{Kiel diagram, showing residuals in \% of $\rho(P)$ at a pressure that is $10^{1.9}$ higher than at the density jump between interpolated models and the corresponding atmospheres in the Stagger-grid at solar metallicity. Models at the boundary of the grid, for which no residuals could be computed, are marked with crosses. The dashed lines show the evolutionary paths of stars, whose masses are given on the plot in units of the solar mass.
}
\label{fig:StaggerError}
\end{figure}

The errors introduced in $\rho(P)$ by interpolation lie below $10\,\%$, for all but one model at high temperature, but are generally of the order of a few percent (cf.~Fig.~\ref{fig:StaggerError}). Atmospheres at the boundary of the grid were not reconstructed in this test. The computed residuals constitute an upper bound for the expected error.

\citet{Magic2016a} presents another interpolation scheme, based on the generic form of the normalized entropy stratification. Comparing our results qualitatively with the published test cases \citep[cf.~Fig.~6 in][]{ Magic2016a}, we conclude that our simple interpolation scheme is able to reconstruct the correct stratifications with similar or even higher accuracy, throughout the ($T_\mathrm{eff},\log g$)-plane.

\begin{figure}
\centering
\includegraphics[width=\linewidth]{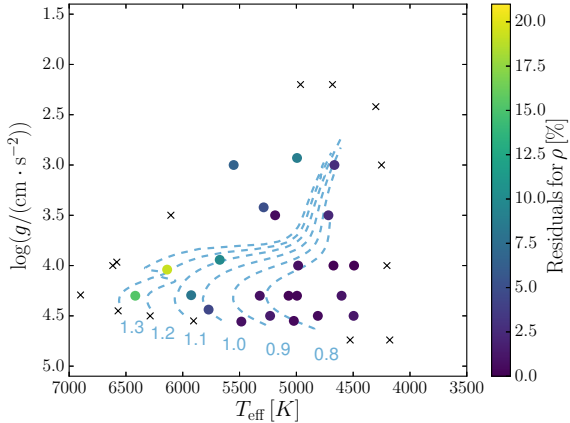}
\caption{Kiel diagram corresponding to Fig.~\ref{fig:StaggerError} but for the Trampedach-grid.
}
\label{fig:TrampedachError}
\end{figure}

Having established that our interpolation scheme yields reasonably accurate results in the case of the regular Stagger-grid, we now take a closer look at the irregular Trampedach-grid. Here we find the accuracy to be strongly affected by the sampling rate. In regions with a high sampling rate, we are able to reconstruct the stratifications within $1\,\%$, and acceptable accuracy is obtained for most main sequence stars. In regions with a low sampling rate, however, the errors may reach $20\,\%$ --- the highest error is again reached for main sequence stars at high temperatures. Fig.~\ref{fig:TrampedachError} shows the error in $\rho(P)$ across the ($T_\mathrm{eff},\log g$)-plane and should be compared to Fig.~\ref{fig:StaggerError}.

As regards the solar atmosphere in the Trampedach-grid, we are able to reproduce $\rho(P)$ only within $5\,\%$, although the sampling is higher in the vicinity of the solar atmosphere than in the corresponding region of the Stagger-grid. We explore the implications of this in Subsection~\ref{subsec:IntTest}.

We note that the stratification of some quantities in the $\langle 3\mathrm{D} \rangle$-atmospheres in the Trampedach-grid shows ripples. In order to remove such ripples, we smooth the interpolated atmospheres before patching, when dealing with the Trampedach-grid, although we found that these ripples have a negligible effect on the model frequencies.

Having established the presented interpolation scheme, it is straight-forward to interpolate in metallicity as well. As this is only possible for the Stagger-grid, and as we only present models that allow for a comparison with atmospheres from the Trampedach-grid, we do not elaborate on this point but leave this issue to be discussed in an upcoming paper.

\section{The Solar Case} \label{sec:solar}

Having summarized the main aspects of post-evolutionary patching above, we proceed to an investigation of how the seismic predictions are affected by the implementation of our procedure. We base our investigation on the Sun, as very detailed observational constraints exist for our host star, whose interior structure is well understood.

In order to perform this asteroseismic analyses, we have computed adiabatic frequencies of stellar p-mode oscillations for patched stellar models, using the Aarhus adiabatic oscillations package, \textsc{adipls} \citep{jcd2008b}. When presenting frequency shifts, $\delta \nu_{n\ell}$, we only include radial modes ($\ell = 0$). There is hence no need to scale each frequency shift with the ratio of the corresponding mode mass and the mode mass of a radial mode with the same frequency.

\subsection{Patched Solar Models} \label{subsec:PatchedSun}

To facilitate an easy comparison with other authors, we have patched Model~S, a standard solar model presented by \citet{jcd1996}. For these PMs, we have computed frequency differences, $\delta \nu_{n\ell}$, between the model frequencies and observations of radial modes ($\ell=0$), stemming from the Birmingham Solar Oscillation Network, BiSON \citep{Broomhall2009, Davies2014}.

For a direct comparison with \citet{Ball2016}, we have employed the same MURaM \citep{Voegler2003,Voegler2005} radiative magnetohydrodynamic (MHD) simulation for the solar atmosphere from \citet{Beeck2013}, averaged over surfaces of constant geometric depth and time.

\begin{figure}
\centering
\includegraphics[width=\linewidth]{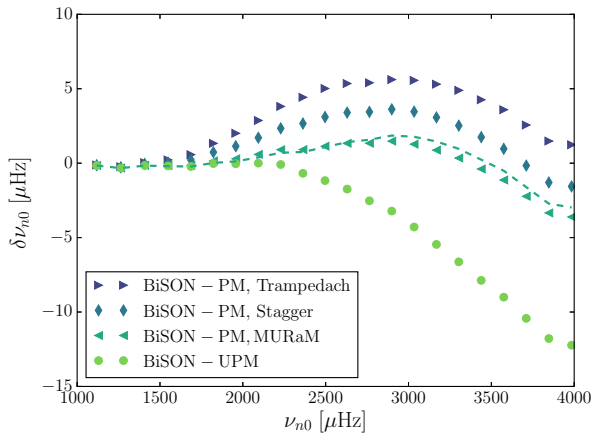}
\caption{Frequency differences between BiSON observations and the un-patched Model~S (UPM) as well as three patched models (PMs), for $\ell=0$. The solar Stagger-grid atmosphere, the solar Trampedach-grid atmosphere and the MURaM simulation were used, respectively. For all models, a weighted combination of $P$ and $\rho$ served as the patching quantity. The patching points are chosen within the adiabatic regions, at the depth, where the closest match in the ($P,\rho$)-plane between the interior and atmosphere is found. The markers are larger than the observational errorbars. The dashed line shows the results presented by \citet{Ball2016} for comparison.
}
\label{fig:DiffAtm}
\end{figure}

In accordance with \citet{Ball2016}, we use $P$ and $\rho$ as patching quantities. Our approach therefore differs from that by \cite{Ball2016} only in minor details. Fig.~\ref{fig:DiffAtm} shows the frequency difference between BiSON data and the model frequencies of the PM. The resulting oscillation frequencies are in good agreement with the results published by \citet{Ball2016}, and we were thus able to reduce the frequency difference between model frequencies and observations from roughly $12\,\mu\mathrm{Hz}$ to $4\,\mu\mathrm{Hz}$. As also stressed in the quoted paper, this is overall consistent with the findings of other authors, including \citet{Piau2014}, who present patched solar models, in which they adopt the average profile of the temperature gradient, $P$, $\rho$, $\Gamma_1$, and the Brunt-V{\"a}is{\"a}l{\"a} frequency from hydrodynamic and magnetohydrodynamic 3D simulations. \citet{Piau2014} have decreased the radius of their UPM, in order to correct for the fact that the PM is slightly larger than the UPM, since $\langle 3\mathrm{D} \rangle$-atmospheres are larger than their 1D counterparts, due to turbulent pressure. Like \citet{Ball2016} and \citet{Magic2016b}, we do not take this effect into account, as we fix the interior structure by Model~S, meaning that the radius of our patched solar models are roughly $100\,$km larger than the constraints on the radius of Model~S.

We have also constructed PMs based on Model~S and the solar atmospheres in the Stagger- and Trampedach-grids. The resulting model frequencies are lower by $2\,\mu\mathrm{Hz}$ and $5\,\mu\mathrm{Hz}$, respectively, at high frequencies, compared to the MURaM solar patched model (cf.~Fig.~\ref{fig:DiffAtm}). We note that these frequency deviations are comparable to the discrepancy between model predictions and observations. As was to be expected, the seismic results are very sensitive to the exact input physics of the atmosphere.

\begin{figure}
\centering
\includegraphics[width=\linewidth]{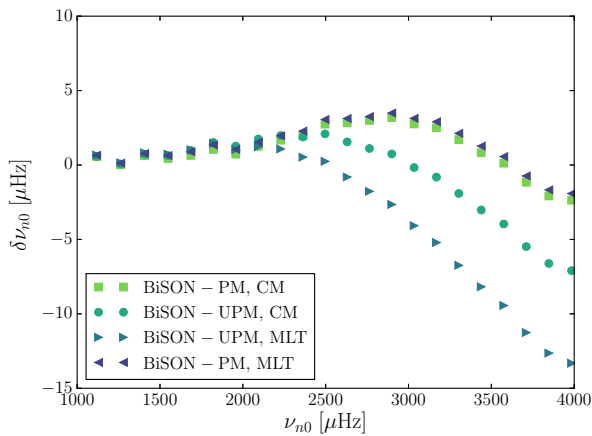}
\caption{Frequency differences between BiSON observations and patched models (PM), for $\ell=0$, based on the solar Stagger-grid model and \textsc{garstec} models with difference convective treatments: mixing length theory (MLT) and the convection theory by Canuto \& Mazzitelli (CM). The models are patched $2.7\,$Mm below the surface, i.e. below a Rosseland optical depth of 1, using $P$ and $\rho$ as patching quantities. 
}
\label{fig:CMMMLT}
\end{figure}

The same holds true for the interior model. Fig.~\ref{fig:CMMMLT} summarizes the results obtained for two different \textsc{garstec} models. Both models use the OPAL-EOS \citep{OPAL2005} and the MHD-EOS in the low-temperature regime, the OPAL opacities \citep{Iglesias1996}, the low-temperature opacities by \citet{Ferguson2005}, and the composition used by \cite{Asplund2009} but differ in their treatment of convection: one uses mixing length theory \citep{Boehm1958}, while the other employes the convection theory by \cite{Canuto1991, Canuto1992} with a solar calibrated mixing length. Although the convection theory by \cite{Canuto1991, Canuto1992} leads to model frequencies (UPM, CM in Fig~\ref{fig:CMMMLT}) that are in better agreement with observations than those using mixing length theory (UPM, MLT in Fig~\ref{fig:CMMMLT}), patching the solar Stagger-grid model to either of the two \textsc{garstec} models yields almost identical results. This means that the patch is performed at a depth, below which the two UPMs are nearly indistinguishable.

As can be seen from Figs.~\ref{fig:DiffAtm} and \ref{fig:CMMMLT}, none of the presented patched solar models actually reproduces the observed BiSON frequencies, as we only take structural effects into account. Modal effects are expected to counteract the structural effects, as stated in the introduction \citep[cf.][]{Houdek2017}. The adiabatic frequencies obtained after solely including structural effects depend on the treatment of turbulent pressure in the frequency calculations \cite[cf.][]{Sonoi2017}. In accordance with \cite{Ball2016}, we assume that the relative Lagrangian perturbation of the turbulent and thermal pressure are equal.

We now take a closer look at the patching procedure. We begin by discussing suitable choices for the patching quantity, as this choice determines both the patching points and the distance of the atmosphere patching point $r_\mathrm{a.p.}(\beta)$ from the stellar centre. The implication is twofold: Firstly, the patching quantity determines the size of the acoustic cavity. Secondly, the stratification of the chosen quantity is smooth, by construction, while discontinuities may occur for the remaining quantities. Since the frequency computation relies on $P(r)$, $\rho(r)$, and $\Gamma_1(r)$, each of these quantities constitutes a suitable choice. Furthermore, some of our stellar models rely on a $T(\tau)$ relation that is extracted from 3D simulations, making the temperature an obvious alternative. The question thus arises as to which degree the seismic results are affected by the choice of the patching quantity.

\begin{figure}
\centering
\includegraphics[width=\linewidth]{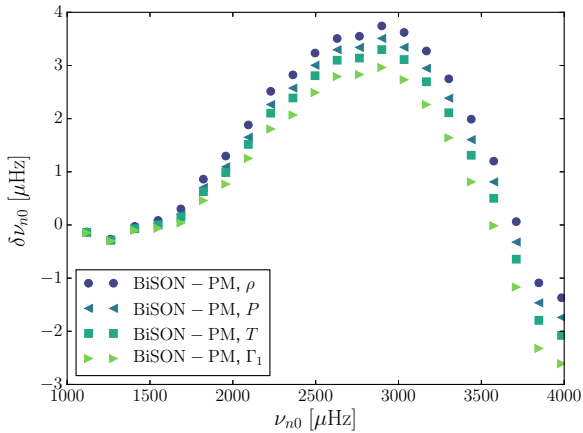}
\caption{Frequency differences between BiSON observations and patched models (PM), for $\ell=0$, based on Model~S and the solar Stagger-grid atmosphere, using different quantities ($P$, $\rho$, $T$, and $\Gamma_1$) for the selection of the interior patching point and $r_\mathrm{a.p.}(\beta)$. All models are patched $2.7\,$Mm below the surface. 
}
\label{fig:SunFreq_diff}
\end{figure}

As can be seen from Fig.~\ref{fig:SunFreq_diff}, PMs, based on different patching quantities, yield the same oscillation frequencies within $1-2\,\mu\mathrm{Hz}$, in the case of the solar Stagger-grid atmosphere. In the case of the mentioned MURaM simulation and the solar Trampedach-grid atmosphere, we find the choice of patching quantities to affect the model frequencies by up to $3\,\mu\mathrm{Hz}$, which is in the same order as the frequency differences presented in Fig.~\ref{fig:DiffAtm} and small compared to the surface effect itself. When dealing with \textit{Kepler} stars in Section~\ref{sec:KeplerPatch}, we will thus present PMs, based on different patching quantities.

As mentioned in Section~\ref{section:methods}, we have also constructed PMs, determining $r_\mathrm{a.p.}(\beta)$, in such a way as to ensure hydrostatic equilibrium to first order, using equation~(\ref{eq:HSEfit}). This procedure results in model frequencies that are consistent with the frequencies obtained, when using $P$ as the patching quantity.

We have varied the patching depth, using different patching quantities: $T$, $P$, $\rho$, $\Gamma_1$, or $c^2 = \Gamma_1 P_\mathrm{gas}/\rho$, where $P_\mathrm{gas}$ denotes the gas pressure. For these patching quantities, we find the model frequencies to be relatively insensitive to the patching depth, if the patching points are situated sufficiently deep within the adiabatic region, independently of the used solar atmosphere.

\citet{Piau2014} have used the optical depth to construct PM. Using this patching quantity in our approach leads to prominent discontinuities in the stratifications of other quantities and yields model frequencies that are rather sensitive to the patching depth. These discontinuities become more prominent with patching depth, which most probably reflects the accumulation of discrepancies that stem from the integration over the opacity, due to discrepancies in the assumed chemical composition. These results reflect the inconsistent physical assumptions that enter the UPM and the $\langle 3\mathrm{D} \rangle$-atmosphere.

Finally, we note that all models presented in Fig.~\ref{fig:DiffAtm}, \ref{fig:CMMMLT}, and \ref{fig:SunFreq_diff} have been computed without correcting for sphericity or discrepancies in $\log g$, in order to facilitate an easy comparison with \citet{Ball2016}. In the case of the shallow solar atmosphere, these corrections have little effect on the overall structure and hence on the oscillation frequencies. For some of the models presented in Subsection~\ref{subsec:MatchingInterior}, on the other hand, these corrections strongly alter the frequencies, due to a sufficiently large mismatch between the gravitational acceleration of the UPM and the gravitational acceleration of the patched atmosphere.

\subsection{Testing the Interpolation Scheme} \label{subsec:IntTest}

In Subsection~\ref{subsec:interpolation}, we have reproduced the solar atmosphere, in both the Stagger-grid and the Trampedach-grid, using our interpolation scheme. In order to assess, whether the accuracy of our scheme is sufficiently high for the purpose of asteroseismic analyses, we have constructed PMs, based on these interpolated solar atmospheres, and compared the seismic results with the model frequencies that were obtained, based on the original solar $\langle \mathrm{3D} \rangle$-atmospheres.

In the case of the Stagger-grid, where the atmosphere structure is reliably reproduced within $2\,\%$ or better, the interpolated solar atmosphere yields the expected model frequencies within $1\,\mu\mathrm{Hz}$, for all frequencies between $1000\,\mu\mathrm{Hz}$ and $4000\,\mu\mathrm{Hz}$, independently of the chosen patching quantity, when performing the patch at the bottom of the atmosphere. The errors introduced by the interpolation scheme are thus in the order of the uncertainty associated with the averaging of the 3D simulations \citep{Ball2016, Magic2016b}.

In the case of the Trampedach-grid, however, a more complex picture emerges. As briefly mentioned in Subsection~\ref{subsec:interpolation}, the interpolation scheme strongly underestimates $\rho(P)$, which affects the selection of the patching points and the computation of $r_\mathrm{a.p.}(\beta)$. Thus, when patching by matching the stratifications in the $(P,\rho)$-plane, as we did in connection with Fig.~\ref{fig:DiffAtm}, the errors in the model frequencies reach $6\,\mu\mathrm{Hz}$ --- better agreement is reached, when using $T$ as the patching quantity.

\subsection{Selecting Atmosphere Parameters} \label{subsec:MatchingInterior}

When constructing PMs with fixed interior structures for \textit{Kepler} stars in Section~\ref{sec:KeplerPatch}, we follow two different approaches for the selection of suitable atmospheres: Firstly, we patch an interpolated atmospheres that have the same $\log g$ and $T_\mathrm{eff}$ as the UPMs. Secondly, we select the global parameters of the atmospheres, based on the requirement to minimize the discontinuities in several quantities near the bottom of the atmospheres, as described in Section~\ref{section:methods}. Both approaches are rather simplistic and have their drawbacks: While the former approach will lead to unphysical discontinuities, minimizing these discontinuities may simply disguise inadequacies of the UPM or the $\langle \mathrm{3D} \rangle$-atmospheres without ensuring a physically correct representation of the stellar structure.

\begin{figure}
\centering
\includegraphics[width=\linewidth]{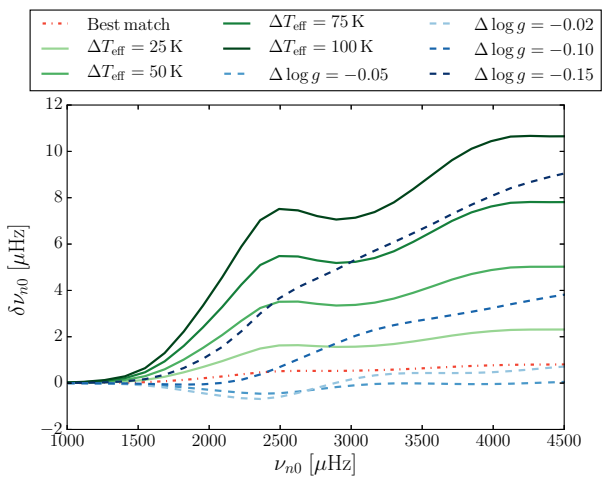}
\caption{Frequency differences obtained from subtracting model frequencies of PMs, based on interpolated atmospheres with different global parameters, from model frequencies, based on the solar Stagger-grid atmosphere. All PMs are constructed, based on Model~S, using $P$ and $\rho$ as patching quantities. We have adjusted $\log g$ to fit the interior model, at the atmosphere patching point, and corrected for sphericity. When varying the effective temperature, $\log g$ has been kept fixed and vice versa. 
}
\label{fig:VaryTg}
\end{figure}

To test, to which extent a mismatch in $T_\mathrm{eff}$ and $\log g$ between the UPM and the patched atmosphere affects the seismic results, we have constructed several PMs, based on Model~S and interpolated atmospheres with different global parameters. The results are presented in Fig.~\ref{fig:VaryTg}. For a mismatch in $T_\mathrm{eff}$ of the order of $100\,$K, we find frequency shifts of the order of $10\,\mu\mathrm{Hz}$, which by far exceeds the error that is expected to arise from the interpolation scheme alone, based on the analysis presented in Subsection~\ref{subsec:IntTest}.

Fig.~\ref{fig:VaryTg} also contains a PM model with the atmosphere that minimizes the discontinuities at the bottom of the atmosphere, i.e. yields the best match to the UPM. In the case of the Stagger-grid, the best match is found at $T_\mathrm{eff}=5778\,\mathrm{K}$, i.e. for an atmosphere, for which the effective temperature lies close to the effective temperature of the UPM. In contrast, we find the effective temperature of the best matching atmosphere to deviate significantly from the effective temperature of the UPM, in the case of the Trampedach-grid. Different selection criteria for the global parameters of the patched atmosphere may hence severely affect the seismic results.

Fig.~\ref{fig:VaryTg} emphasizes the importance of the interpolation scheme presented in Section~\ref{section:methods}: Simply selecting the nearest atmospheres in the grid will severely distort the seismic results and is hence unsuitable for correcting structural effects. Patching the nearest atmosphere in the Stagger-grid, after excluding the solar atmosphere, shifts the model frequencies by up to $19\,\mu\mathrm{Hz}$.

\section{Stars in the \textit{Kepler} Field} \label{sec:KeplerPatch}

We now present an asteroseismic analysis of several stars in the \textit{Kepler} field, employing both UPMs and PMs. We only consider solar-type dwarfs in a limited mass range, guided by the established accuracy of the interpolation scheme in different regions of the ($T_\mathrm{eff},\log g$)-plane. Due to the restrictions imposed by the Trampedach-grid, we selected objects that have metallicities consistent with the assumption of solar metallicity within one standard deviation of the spectroscopic values listed in \citet{Lund2017} and \citet{Aguirre2015}.

Keeping the composition fixed, leaves us with only the mass and evolutionary stage to vary, when selecting the UPM that best fit observations. At this stage, we avoid the additional interpolation in metallicity. We also ignore effects of atomic diffusion, thus assuming that the present surface composition is the initial one.

\subsection{Selecting UPMs} \label{subsec:selectUPM}

To ensure consistency between the interior model and the patched $\langle 3\mathrm{D} \rangle$-atmosphere, we construct UPMs, employing the $T(\tau)$ relation and calibrated $\alpha_\mathrm{MLT}$ \citep{Trampedach2014a, Trampedach2014b} implemented by \citet{Mosumgaard2016} and the composition imposed by \citet{Trampedach2013}. Using \textsc{garstec}, we have constructed a grid of stellar models, computing evolutionary tracks for models with masses between $0.70\,\mathrm{M}_\odot$ and $1.20\,\mathrm{M}_\odot$ in steps of $0.01\,\mathrm{M}_\odot$. This grid covers the parameter range of the 3D simulations. Using \textsc{adipls}, we have then computed model frequencies for all models, for which the large frequency separation, $\Delta \nu$, lies between $60\,\mu\mathrm{Hz}$ and $\mathrm{170\,\mu\mathrm{Hz}}$, which for a $1\,M_\odot$ star covers the evolution from the ZAMS to the onset of the red giant branch. We refer to these as 3D-calibrated models. For the purpose of consistency, these models employ the MHD-EOS at low temperatures and the same abundances as used in the hydrodynamic simulations by \citet{Trampedach2013}. Where the MHD-EOS is not available, the OPAL-EOS is used. We use the same monochromatic low-temperature opacities as presented in \cite{Trampedach2014b} merged with the OP high temperature opacities \citep{Badnell2005}, for the same mixture as in the Trampedach-grid, and conductive opacities by \cite{Cassisi2007}.

Above the photosphere, the structure of the 3D-calibrated UPMs are dictated by the $T(\tau)$ relation derived from the 3D simulations. Below the photosphere, the 3D simulations are taken into account via the variable $\alpha_\mathrm{MLT}$ and adjustments to the radiative gradient. For further details, we refer to \cite{Trampedach2014a,Trampedach2014b} and \cite{Mosumgaard2016}. Nevertheless, the 3D-calibrated UPMs do not perfectly reproduce the $\langle 3\mathrm{D} \rangle$-atmospheres. The 3D-calibrated UPMs have been computed neglecting turbulent pressure, while the turbulent pressure significantly contributes to the total pressure throughout large parts of the patched atmospheres. Disregarding the turbulent pressure also leaves traces in the structure at depths, at which it has become negligible. Furthermore, the models do not take convective back-warming\footnote{Convective temperature fluctuations and the extreme temperature sensitivity of the opacity affect the cooling of the granules and the inter-granular lanes asymmetrically. This results in a warming below the photosphere and an expansion of the atmosphere that is not accounted for by 1D models but found in 3D simulations. We refer to \cite{Trampedach2013, Trampedach2017} for further details.} into account, and the employed mixing length that takes a constant value within each envelope cannot encapsulate the detailed structure of 3D-simulations. Consequently, discontinuities also occur for these models, when patching below the photosphere (cf.~Subsection~\ref{subsec:discussion}).

While the outermost layers of the 3D-calibrated models do not perfectly mimic the underlying 3D simulations, the implementation of the associated $T(\tau)$ relation leads to more physically accurate boundary conditions than the Eddington atmosphere, giving a better description of the transition between optically thin an optically thick regions. Furthermore, the calibrated $\alpha_\mathrm{MLT}$ that is employed by the 3D-calibrated models varies with $T_\mathrm{eff}$ and $\log g$ and gives a more realistic description of convection throughout the Hertzsprung-Russell-diagram than a constant mixing length does. Finally, while the structure of the outermost layers of the 3D-calibrated models does not perfectly match the corresponding $\langle 3\mathrm{D} \rangle$-atmospheres, the implementation by \cite{Mosumgaard2016} guaranties a high level of consistency between the input physics of these models and the input physics of the 3D simulations. 

We also include models \citep[cf.][]{Sahlholdt2017} based on a different grid that employs the OPAL-EOS, the OPAL opacities, the low-temperature opacities by \cite{Ferguson2005}, the conductive opacities by \cite{Cassisi2007}, and the composition \cite{Grevesse1998}, for comparison. These models use conventional Eddington grey atmospheres. This grid includes stars with masses between $0.70\,\mathrm{M}_\odot$ and $1.30\,\mathrm{M}_\odot$ in steps of $0.01\,\mathrm{M}_\odot$, for which the large frequency separation lies between $13\,\mu\mathrm{Hz}$ and $\mathrm{180\,\mu\mathrm{Hz}}$. The metallicity is treated as a free parameter and varied between $-0.50\,$dex and $0.50\,$dex in steps of $0.05\,$dex. We refer to these models as standard input models. The helium mass fraction is varied between 0.25 and 0.30 in steps of 0.002, and the solar calibrated mixing length parameter is set to 1.79.

We compare the obtained model frequencies with the observed frequencies extracted from \textit{Kepler} data, based on a Bayesian Markov Chain Monte Carlo 'peak-bagging' approach \citep[cf.][]{Davies2016, Lund2017}. The best model is selected, based on the individual frequencies after applying the empirical surface correction by \citet{Ball2014}.

For comparison, we have also found the UPM that yields the best fit in frequency ratios: $r_{01}$, $r_{10}$, and $r_{02}$. This approach was originally proposed by \citet{Roxburgh2003}, as these ratios have been shown to be rather insensitive to the structure of the outermost layers \citep{Oti2005}:
\begin{equation}
r_{01}(n) = \frac{\nu_{n-1,0}-4\nu_{n-1,1} + 6 \nu_{n,0} - 4\nu_{n,1} + \nu_{n+1,0}}{8(\nu_{n,1}-\nu_{n-1,1})},
\end{equation}
\begin{equation}
r_{10}(n) = \frac{-\nu_{n-1,1} + 4\nu_{n,0} - 6\nu_{n,1} + 4\nu_{n+1,0} - \nu_{n+1,1}}{8(\nu_{n+1,0}-\nu_{n,0})},
\end{equation}
\begin{equation}
r_{02}(n) = \frac{\nu_{n,0}-\nu_{n-1,2}}{\nu_{n,1}-\nu_{n-1,1}},
\end{equation}
where $\nu_{n,\ell}$ denotes the individual oscillation frequency of radial order $n$ and degree $\ell$. 

We determine the stellar parameters by choosing the model that gives the maximal likelihood, based on the listed frequency ratios or the individual frequencies, using the \textbf{BA}yesian \textbf{ST}ellar \textbf{A}lgorithm, \textsc{basta} \citep{Aguirre2015}.

\subsection{Selecting Atmospheres}

\begin{table*}
	\centering
	\caption{Parameters for UPMs and PMs of \textit{Kepler} stars. The indices refer to the probability distribution (P), from which the UMPs are drawn, the UPMs that lead to the best fit (U), the PMs, based on the Stagger-grid (S), and the PMs, based on the Trampedach-grid (T), respectively. The index 'con' indicates spectroscopic and asteroseismic constraints on $T_\mathrm{eff}$ and $\log g$ that have been adopted from either the LEGACY (1) or KAGES (2) sample, which is specified in the last column. Specific references to papers are given in the text. The uncertainties given for the UPM denote $68.3\,\%$ credibility intervals evaluated by \textsc{BASTA}, based on the employed grid of UPMs, fitting individual frequencies. The asterisk denotes standard input models, while the rest are 3D-calibrated models.}
	\label{garstecFeH0}
	\begin{tabular}{lccccccccccccccccccccc} 
		\hline
		KIC  & $T_\mathrm{eff,con}$ & $\log g_{\mathrm{con}}$ & $M_\mathrm{P}$ & $\mathrm{[Fe/H]}$ & $T_\mathrm{eff,P}$ & $\log g_{\mathrm{P}}$ & $T_\mathrm{eff,U}$ & $\log g_{\mathrm{U}}$ & $T_\mathrm{eff,S}$ & $T_\mathrm{eff,T}$ & Sample \\
		& $(\mathrm{K})$ & $(\mathrm{K})$ & $\mathrm{(cgs;\,dex)}$ & $(\mathrm{M}_\odot)$ & $\mathrm{(dex)}$ & $(\mathrm{K})$ & $\mathrm{(cgs;\,dex)}$ & $(\mathrm{K})$ & $\mathrm{(cgs;\,dex)}$ & $(\mathrm{K})$ & \\		
		\hline
        9025370* &  $5270\pm180$ & $4.423^{+0.004}_{-0.007}$ & $1.016^{+0.007}_{-0.007}$ & $0.06^{+0.09}_{-0.07}$ & $5758^{+71}_{-61}$ & $4.430^{+0.003}_{-0.003}$ & 5771 & 4.430 & 5773 & 5729 & 1 \\[6pt]
		9025370 & $5270\pm180$ & $4.423^{+0.004}_{-0.007}$ & $0.988^{+0.019}_{-0.008}$ & 0 & $5905^{+38}_{-19}$ & $4.428^{+0.003}_{-0.003}$ & 5908 & 4.427 & 5761 & 5764 &  1 \\[6pt]
		9955598* & $5457\pm77$ & $4.497^{+0.005}_{-0.007}$ & $0.900^{+0.01}_{-0.01}$ & $0.07^{+0.07}_{-0.08}$ & $5370^{+61}_{-61}$ & $4.497^{+0.002}_{-0.002}$ & 5396 & 4.497 & 5506 & 5417 &  1  \\[6pt]
		9955598 & $5460\pm75$ & $4.495^{+0.002}_{-0.002}$ & $0.884^{+0.003}_{-0.005}$ & 0 & $5599^{+19}_{-26}$ & $4.496^{+0.002}_{-0.001}$ & 5596 & 4.496 &  5506 & 5445 &  2  \\[6pt]
		11133306* & $5982\pm82$ & $4.314^{+0.004}_{-0.007}$ & $1.05^{+0.04}_{-0.04}$ & $0.0^{+0.1}_{-0.1}$ & $5956^{+68}_{-78}$ & $4.313^{+0.007}_{-0.007}$ & 5960 & 4.313 & 5991 & 5904 & 2 \\[6pt]
		11133306 & $5982\pm82$ & $4.314^{+0.004}_{-0.007}$ & $1.01^{+0.03}_{-0.02}$  & 0 & $6010^{+70}_{-70}$ & $4.307^{+0.006}_{-0.008}$ & 5998 & 4.306 & 5989 & 5870 &   2 \\[6pt]
        11772920* & $5180\pm180$ & $4.500^{+0.008}_{-0.005}$ & $0.840^{+0.01}_{-0.02}$ & $-0.11^{+0.08}_{-0.06}$ & $5345^{+70}_{-60}$ & $4.503^{+0.003}_{-0.003} $ & 5354 & 4.502 & 5506 & 5330 &   1 \\[6 pt]
		11772920 & $5180\pm180$ & $4.500^{+0.008}_{-0.005}$ & $0.854^{+0.003}_{-0.004}$ & 0 & $5476^{+28}_{-21}$ & $4.506^{+0.002}_{-0.002}$ & 5479 & 4.505 & 5506 & 5313 &   1 \\
		\hline
	\end{tabular}
\end{table*}

The stellar parameters for the UPMs, selected as described above to model the four \textit{Kepler} stars, are listed in Table~\ref{garstecFeH0} alongside observational constraints and the global parameters for the interpolated $\langle 3\mathrm{D} \rangle$-atmospheres that minimize the discontinuities, at the bottom of the atmosphere, as described in Section~ \ref{section:methods}. Furthermore, Table~\ref{garstecFeH0} summarizes the values and uncertainties of the stellar mass, $T_\mathrm{eff}$ and $\log g$ as obtained from \textsc{basta}, i.e. key values of the mapped probability distributions. Note that the listed median does not correspond to the best fitting model.

The observational constraints on the effective temperature and metallicities for stars in the LEGACY dwarfs sample~I are derived spectroscopically and adopted from Table~1 in \citet{Lund2017}; in the case of KIC~9025370 and KIC~11772920, the spectroscopic values are from \citet{Pinsonneault2014}. The constraints on the gravitational acceleration are derived asteroseismically and are taken from Table~3 in \citet{Aguirre2017}. In the case of the stars from the KAGES sample, the spectroscopic values for the effective temperatures as well as the asteroseismic constraints on $\log g$ are adopted from Table~3 in \citet{Aguirre2015}. 

For both samples, the listed constraints on $\log g$ are obtained from modelling the individual frequencies, using \textsc{basta}. One may object to using the outcome of stellar modelling, when considering models with different input physics. However, as discussed in Section~\ref{section:methods}, we only employ these constraints in order to restrict the region of interest, when looking for the best match in the ($T_\mathrm{eff},\log g$)-plane.

As can be seen from Table~\ref{garstecFeH0}, the metallicities of the best fitting standard input UPMs are in agreement with the assumption of solar metallicity, the evaluated $\log g$ are in good agreement with the findings of \citet{Aguirre2015, Aguirre2017}, and the masses of the standard input UPMs are in very good agreement with the masses of the 3D-calibrated UPMs --- and with the findings of \cite{Aguirre2017}.

For all four \textit{Kepler} stars, the effective temperature of the standard input UPMs are lower and in better agreement with the spectroscopic constraints than their 3D-calibrated counterparts are.

\subsection{Asteroseismic Analyses of Patched Models}

For all eight UPMs listed in Table~\ref{garstecFeH0}, we have constructed 6 patched models: For each grid, we have constructed 2 PMs, employing either $T$ or a combination of $P$ and $\rho$ as patching quantities and using atmospheres, for which $T_\mathrm{eff}$ and $\log g$ match the parameters of the UPM. Furthermore, we have constructed 1 PM for each grid, using $P$ and $\rho$ as patching quantities, and selecting the atmosphere, in such a way as to minimize the discontinuities between the interior model and the $\langle \mathrm{3D} \rangle$-atmosphere, near the bottom of the atmosphere. These patching criteria have been selected based on our investigation of the solar case, from which we conclude that the choice of patching quantity and the selection of the patched atmosphere have the largest effect on the seismic results.

\subsubsection{KIC 9025370}

\begin{figure}
\centering
\includegraphics[width=\linewidth]{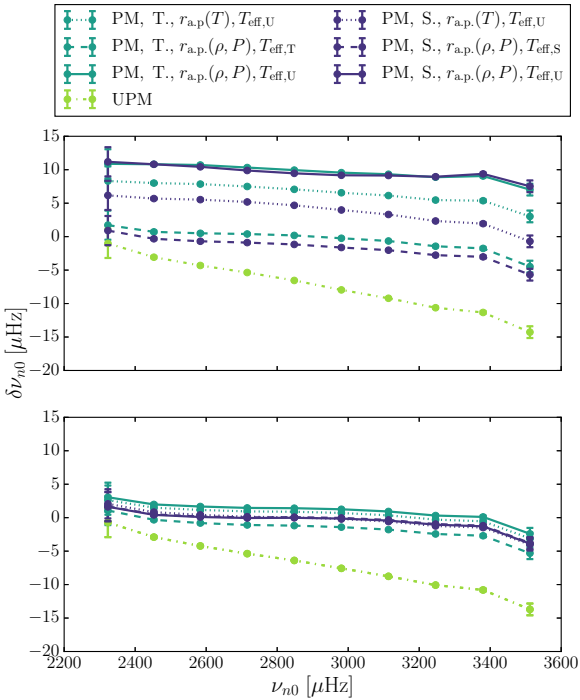}
\caption{Frequency differences between observations and 2 UPM as well as 12 PMs, for KIC~9025370, based on the LEGACY frequencies. \textbf{Upper panel:} 3D-calibrated UPM. \textbf{Lower panel:} standard input UPM. The plot includes PMs, based on interpolated Trampedach-grid (T.) and Stagger-grid (S.) atmospheres with the same $T_\mathrm{eff}$ ($T_\mathrm{eff,U}$) as the UPM, as well as PMs, based on atmospheres with $T_\mathrm{eff,T}$ and $T_\mathrm{eff,S}$ (cf.~Table~\ref{garstecFeH0}). Different patching quantities are used: either a combination of $P$ and $\rho$ or $T$. In all cases, the patching points are chosen, at the depth, at which the closest match in the $(P,\rho)$- or $(P,T)$-plane between the interior and the atmosphere is found.
}
\label{fig:KIC9025370}
\end{figure}

The upper and lower panel of Fig.~\ref{fig:KIC9025370} show a comparison between observations and model frequencies for the UPM and the associated 6 PMs of KIC~9025370. We have selected the UPM, based on the individual frequencies, and find that an analysis, relying on the frequency ratios, leads to similar results but depends slightly on the restrictions on $\Delta \nu$.

We note that the effective temperature of the 3D-calibrated UPM does not lie within three standard deviations of the spectroscopic value and that the effective temperature of the standard input model likewise lies above the spectroscopic value. This may suggest that the spectroscopic constraints underestimate the effective temperature in the case of KIC~9025370. 

\subsubsection{KIC 9955598}

\begin{figure}
\centering
\includegraphics[width=\linewidth]{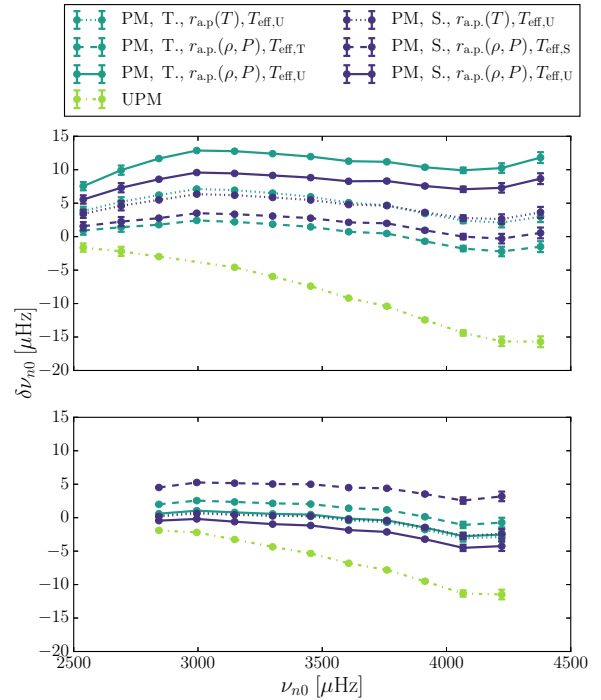}
\caption{As Fig.~\ref{fig:KIC9025370} but for KIC~9955598. \textbf{Upper panel:} 3D-calibrated UPM, using KAGES frequencies. \textbf{Lower panel:} standard input UPM, using LEGACY frequencies. The associated effective temperatures are listed in Table~\ref{garstecFeH0}.
}
\label{fig:KIC9955598}
\end{figure}

Fig.~\ref{fig:KIC9955598} summarizes the computed model frequencies for KIC~9955598 that appears in both samples. For comparison we have selected the 3D-calibrated UPM following different approaches. Similar best fitting UPMs are found, when fitting frequency ratios and individual frequencies from either sample. 

\subsubsection{KIC 11133306}

\begin{figure}
\centering
\includegraphics[width=\linewidth]{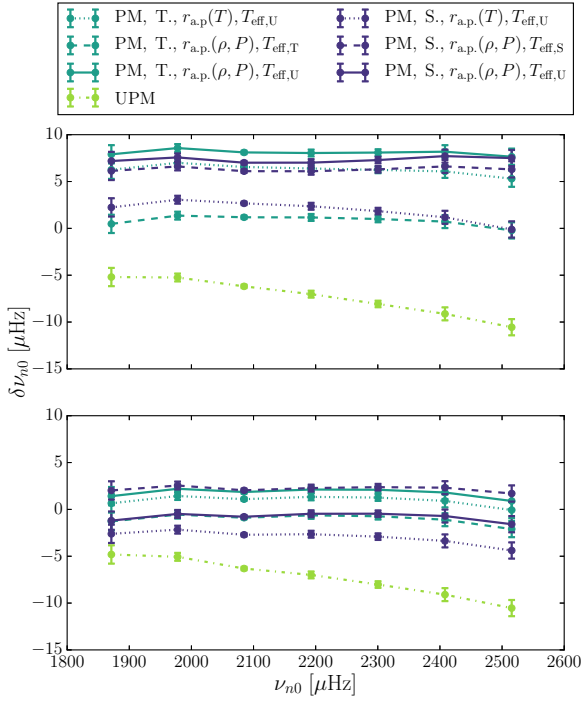}
\caption{As Fig.~\ref{fig:KIC9025370} but for KIC~11133306, based on the KAGES sample frequencies. The associated effective temperatures are listed in Table~\ref{garstecFeH0}.
}
\label{fig:KIC11133306}
\end{figure}

Fig.~\ref{fig:KIC11133306} summarizes the seismic results obtained for KIC~11133306. When constructing PMs for KIC~11133306, based on either UPM and the Stagger-grid, minimization of the discontinuities near the bottom of the patched atmosphere does not lead to a unique solution, within the spectroscopic and asteroseismic constraints. On the contrary, two distinct local minima are found in the investigated region of the $(T_\mathrm{eff},\log g)$-plane. We chose the atmosphere, whose effective temperature lies closest to the effective temperature of the UPM. This underlines the necessity of establishing a rigid scheme for the selection of the $\langle 3\mathrm{D}\rangle$-atmosphere, when constructing PMs, based on a given UPM. 

\subsubsection{KIC 11772920}

\begin{figure}
\centering
\includegraphics[width=\linewidth]{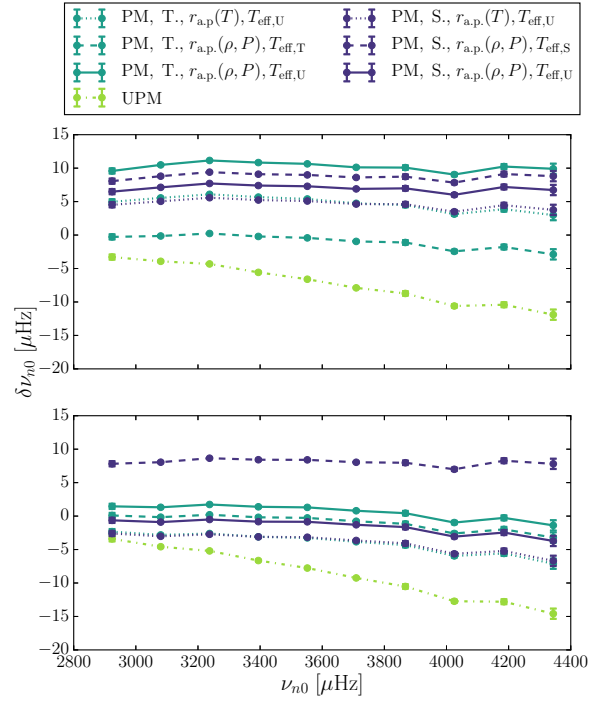}
\caption{As Fig.~\ref{fig:KIC9025370} but for KIC~11772920, based on the LEGACY sample frequencies.The associated effective temperatures are listed in Table~\ref{garstecFeH0}.
}
\label{fig:KIC11772920}
\end{figure}

Fig.~\ref{fig:KIC11772920} summarizes the results obtained for KIC~11772920. As for case of KIC~11133306, the suggested minimization of discontinuities does not lead to a unique solution, in the case of the Stagger-grid, for either UPM. We thus find local minima at $5043\,\mathrm{K}$ and at $5506\,\mathrm{K}$. However, when constructing PMs, based on the colder atmosphere, we obtained models that are smaller than the corresponding UPM, which is inconsistent with our other patched models. Consequently, we discard this solution and set for the warmer minimum.

\subsection{Discussion} \label{subsec:discussion}

All in all, the obtained model frequencies are in good agreement with the qualitative conclusions drawn from the solar case: patching 3D atmospheres reduces the model frequencies, and using different atmosphere grids, interior models or patching quantities leads to frequency difference that are of order of a few microhertz. These are small compared to the surface effect itself but substantial compared to the residual discrepancy between model frequencies and observations.

While the surface effect increases with frequency for the UPMs, the discrepancies between observations and the model frequencies for all 48 PMs are not strongly frequency dependent. Again, we emphasize that the PMs are not expected to reproduce the observations, since we have not taken modal effects into account. According to \cite{Houdek2017}, disregarding modal effects is rather expected to lead to PMs that underestimate and hence over-correct the eigenfrequencies. While most of the 3D-calibrated PMs over-correct the model frequencies, roughly half of standard input PMs do not. Based on the considerations listed above, the PMs that already under-correct the frequencies are not expected to correctly reproduce the structure of the considered \textit{Kepler} stars, if the modal effects are added. In this respect, the 3D-calibrated models perform much better than their standard input counterparts.

The discrepancy in effective temperature between the employed UPM and the atmosphere that minimizes the discontinuities are generally found to be larger for the 3D-calibrated models (cf.~Table~\ref{garstecFeH0}). Just as in the solar case, such high discrepancies in the effective temperature strongly affect the evaluated surface correction. Consequently, the frequency differences between the 3D-calibrated PMs are larger than in the case of the standard input models. This may reflect discrepancies between the 3D-calibrated models and the 3D-atmospheres but may also issue from the fact that the approaches for the selection of the global parameters of the patched atmospheres are rather simplistic.

Each of the four 3D-calibrated \textit{Kepler} models has a higher effective temperature than the respective standard input models do. This, at least partly, reflects the fact that we kept the metallicity fixed, when determining the best fitting 3D-calibrated models, while the metallicity entered the analysis as a variable, in the case of the standard input models. Moreover, in contrast to the standard input models, the 3D-calibrated models have been computed neglecting atomic diffusion, due to the restrictions imposed by the Trampedach-grid. According to \cite{Aguirre2015}, the neglect of diffusion affects the global parameters of the best fitting UPMs.

The eigenfrequencies of the 3D-calibrated models are more sensitive to the choice of the patching quantity than the standard input models are. A larger sensitivity to the choice of the patching quantity implies a larger mismatch between the patched atmosphere and the structure of the UPM at the patching point, i.e. larger discontinuities. However, since the frequency shifts that are introduced by changing the patching quantity are of the order of a few microhertz, several factors may contribute to this behaviour: Firstly, based on our analysis of the solar case, we conclude that the sensitivity to the choice of the patching quantity, at least partly, reflects the accuracy of the interpolation scheme. Secondly, we note that a direct comparison between the panels of Figs~\ref{fig:KIC9025370}-\ref{fig:KIC11772920} is misleading, since the global parameters of the 3D-calibrated models and their standard input counterparts are not the same. Indeed, for all four \textit{Kepler} stars, the best fitting 3D-calibrated UPM has a higher effective temperature than the associated best fitting standard input model, and, as discussed in Section~\ref{section:methods}, the interpolation error increases with increasing effective temperature (cf.~Fig.~\ref{fig:TrampedachError}). Thirdly, a direct comparison between the panels is misleading, since the 3D-calibrated models use different equations of state, different opacity tables, and a different mixture than their standard input counterparts.

\begin{figure}
\centering
\includegraphics[width=\linewidth]{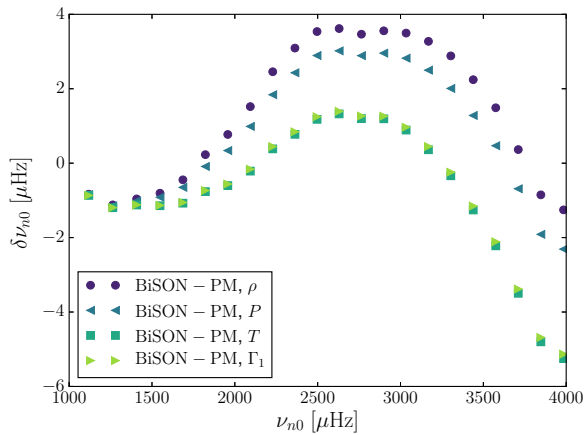}
\caption{As Fig.~\ref{fig:SunFreq_diff} but for a 3D-calibrated solar model and the solar Trampedach-grid atmosphere. All models are patched $2.5\,$Mm below the surface.}
\label{fig:Sol3D}
\end{figure}

As elaborated upon in Subsection~\ref{subsec:selectUPM}, the outermost layers of the 3D-calibrated models do not perfectly mimic the underlying 3D simulations, which contributes to the sensitivity to the choice of the patching quantity.
In order to access, how well the implementation of the $T(\tau)$ relation and the calibrated $\alpha_{MLT}$ reproduces the structure of the Trampedach-grid atmospheres, we present PMs, based on a 3D-calibrated solar UPM, in Fig.~\ref{fig:Sol3D}. The 3D-calibrated solar UPM has the same effective temperature as the solar Trampedach-grid atmosphere. Consequently, no interpolation has been performed. As can be seen from the figure, we find the choice of the patching quantity to affect the eigenfrequencies by up to $4\,\mu\mathrm{Hz}$, i.e. by roughly the same amount, as we found for Model S, when patching the same solar atmosphere ($3\,\mu\mathrm{Hz}$, cf.~Subsection~\ref{subsec:PatchedSun}). This frequency difference reflects the structural change that arises from a shift in the distance of the patching point from the solar centre: $r_\mathrm{a.p.}(T)$ is roughly $40\,\mathrm{km}$ less than $r_\mathrm{a.p.}(P)$, as the temperature of the atmosphere is roughly $1\,\%$ higher at the base than the temperature of the UPM at the same pressure. In accordance with our study of Model S, we hence find the eigenfrequencies to be very sensitive to changes in the structure of the PM. 

For comparison, we have performed yet another solar calibration with the same abundances, equations of state, opacity tables, and effective temperature as were employed for the solar model in Fig.~\ref{fig:Sol3D} but with a gray Eddington atmosphere and a constant mixing length along the evolutionary track. We find this solar model to show the same sensitivity to the patching criteria as the model in Fig.~\ref{fig:Sol3D} does. Changing the boundary conditions and the mixing length merely increases the model frequencies by less than $1\,\mu\mathrm{Hz}$ for all patching quantities. This finding is in good agreement with the results presented in Figure~\ref{fig:CMMMLT}, showing that the boundary conditions have little influence on the interior structure sufficiently deep within the star. 

To facilitate an easy comparison between Fig.~\ref{fig:Sol3D} and Figs~\ref{fig:DiffAtm}-\ref{fig:SunFreq_diff}, we have not corrected for sphericity or adjusted $\log g$ to fit the interior model, at the atmosphere patching point. Applying such corrections increases the model frequencies of the presented solar model by up to $1\,\mu\mathrm{Hz}$.

\section{Conclusions} \label{sec:Conclusion}

In this paper, we present an analysis of patched stellar models (PMs), investigating several new aspects, in order to establish the necessary tools for an improvement of theoretical predictions and a more realistic treatment of pulsation properties in near-surface layers. We present a new method for the interpolation of the mean structures of 3D simulations. Our scheme can be directly employed in asteroseismic analyses in stead of solely allowing for the calibration of relations for surface corrections.

We have tested our interpolation scheme by reconstructing existing atmospheres. At least in the case of cold main sequence stars, we find our method to accurately reproduce the correct atmosphere structure. Concerning warmer stars and other evolutionary stages, the interpolated structures are not sufficiently accurate for the purpose of asteroseismology. We attribute this to the low sampling of current atmosphere grids, and hence our results call for a refinement of these grids.

Having established a robust interpolation scheme, we investigate, how a mismatch between the effective temperature and gravitational acceleration of the un-patched model (UPM) and the patched mean 3D atmosphere affects the seismic results. The fact that we consistently find even small mismatches to have a non-negligible effect on the model is rather crucial. Thus, in this paper, we present PMs that either have the same effective temperature as the UPM or are selected in such a way as to minimize the discontinuities in several quantities, near the bottom of the atmosphere. Although both approaches are rather simplistic, they illustrate the importance of establishing a rigid matchmaking scheme, as the two approaches lead to different model frequencies, in several cases.

We also investigate how different patching criteria affect the seismic results, based on patched solar models. These criteria include the depth, at which the patch is performed, and the quantities, based on which the patching points are selected, and based on which the distance between the stellar centre and the lowermost point in the patched mean 3D atmosphere is established. While the model frequencies are mostly unaffected by the patching depth, if the patch is performed sufficiently deep within the adiabatic region, the choice of patching quantity may shift the model frequencies by a few microhertz. These shifts are small compared to the surface effect itself but comparable to the residual discrepancy between model frequencies and observations. Furthermore, the model frequencies are likewise sensitive to the input physics, and hence to the employed atmosphere grid.

As regards our patched solar models, the resulting model frequencies are in very good agreement with an analysis presented by \citet{Ball2016}. Moreover, our results are consistent with the observed solar oscillation frequencies, considering that modal effects have not been taken into account.

Regarding the sensitivity of the model frequencies on the employed patching quantities and on the input physics, our analysis of \textit{Kepler} stars leads to the same qualitative conclusions as in the solar case. Just as for the Sun, patching a mean 3D atmosphere lowers the model frequencies. The remaining frequency differences between model frequencies and observations are less frequency dependent than in the case of the UPM.

Based on the presented results, we are hence able to identify several steps in the post-evolutionary patching procedure, for which different seemingly equally valid choices moderately or severely affect the outcome of asteroseimic analyses. These ambiguities in the eigenfrequencies should thus be taken into account, when using post-evolutionary patching to correct for surface effects --- say, in order to calibrate relations for surface corrections. Furthermore, the sensitivity of the eigenfrequencies to the patching criteria limits the improvement that can be achieved from patching 1D stellar models.

All in all, the present paper clearly underlines the potential of post-evolutionary patching, showing how the implementation of information from 3D hydrodynamic simulations can be used to improve 1D stellar models. Based on our patching procedure, we are thus partly able to correct for surface effects, obtaining results that are in very good agreement with qualitative expectations. Apart from taking modal effects into account, further improvements include an equally robust implementation the results from 3D simulations, throughout the stellar evolution. The implementation of the $T(\tau)$ relation and the calibrated $\alpha_\mathrm{MLT}$ from 3D simulations \citep{Mosumgaard2016} constitutes an essential improvement along these lines --- how models are affected by this implementation alone is investigated in an independent paper (Mosumgaard et al., in prep.). However, the outermost layers of the resulting models do still not perfectly match the corresponding $\langle 3\mathrm{D} \rangle$-atmospheres (cf.~Subsection~\ref{subsec:selectUPM}). For the purpose of asteroseismology, further improvement of 1D models is hence required. The patching method presented in this paper allows to correct for these structural inadequacies for any model within the parameter space mapped by 3D simulations. This is a novel and significant improvement compared to previous papers that have yet avoided interpolation of 3D simulations in connection with the construction of patched models. The next step is to include interpolated mean 3D structures on the fly instead of at a certain evolutionary stage. While further improvement will ideally ensure the correct mean structure of 1D models, making patching obsolete, our method is currently the most apposite approach.

\section*{Acknowledgements}

We record our gratitude to W.~Ball, J.~Christensen-Dalsgaard, R.~Collet, Z.~Magic, and R.~Trampedach for the collaboration, their many insights, their useful input, and fruitful discussions. Funding for the Stellar Astrophysics Centre is provided by the Danish National Research Foundation (Grant DNRF106). We also acknowledge support from Villum Fonden (research grant 10118).












\bsp	
\label{lastpage}
\end{document}